\documentclass{article}
\usepackage{geometry}
\usepackage{graphicx} 
\usepackage{amssymb}
\usepackage{amsfonts}
\usepackage{amsmath}
\usepackage[sorting=none]{biblatex}
\bibliography{bibli-02}
\title{3HDM-multidim-log}
\author{Joris Vergeest}

\begin{document}
\begin{centering}
{\Large \bf Lepton masses and mixing in a three-Higgs doublet model}\\
\vskip 0.2cm
{\large Joris Vergeest, Bartosz Dziewit, Piotr Chaber\\}
University of Silesia, Katowice, Poland\\
\vskip 0.2cm
{\large Marek Zra\l ek}\\
Humanitas University, Sosnowiec, Poland
\vskip 1.2cm
{\Large \textbf Abstract}
\end{centering}
\vskip 0.2cm
In the three-Higgs doublet model (3HDM) frame, we search for discrete flavour symmetries that give relations among the lepton masses and their mixing angles. We explore discrete non-Abelian groups of order less than 1035, treating neutrinos as Majorana or Dirac particles. Despite the free dynamic parameters available in the model, none of the groups fully predicts the lepton data. However, some of the scanned groups provide either the correct neutrino masses and mixing angles or the correct masses of the charged leptons. $\Delta(96)$ is the smallest group compatible with the experimental data of neutrino masses and PMNS mixing. $S_4$ is an approximate symmetry of Dirac neutrino mixing, with parameters staying about $3\sigma$ apart from the measured $\theta_{12}$, $\theta_{23}$, $\theta_{13}$.
\vskip 1cm
\section{Introduction}
Within the Standard Model, despite its success, still there are basic open questions. For example, the question of the origin of the three fermion generations, with such diverse mass hierarchies and radically differing mixing patterns of leptons, compared to quarks is unsolved \cite{Harari:1977kv,IBE2016365}. In the lepton sector, which is the focus of this work, various attempts have been made to devise a theory to predict the neutrino masses and the lepton mixing angles (e.g. see \cite{King:2014nza, Qian:2015waa, Frampton:1994rk}). As the experimental data on these quantities have gained precision in the recent years (e.g. see \cite{Aker:2022ijt, T2K:2019bcf, RENO:2018dro}), it is getting increasingly hard to explain them satisfactorily. A common approach to this problem is to look for lepton flavor symmetries of the interaction Lagrangian, which may pin down mass values and the neutrino mixing angles, and if so, to determine whether these are (partially) consistent  with the experimental data. It is well known that an exact nontrivial flavor symmetry in the lepton sector does not exist after EWSB, which follows from the fact that the lepton masses are distinct \cite{Feruglio:2019ybq}. Due to Schur's first lemma, the acting of two inequivalent representations in flavor space (one on the lepton doublets, one on the lepton singlets) implies that any mass matrix is either proportional to the unit matrix or vanishes. One strategy to avoid the above problem is to break a remnant flavor symmetry explicitly. Then the charged lepton mass matrix and the neutrino mass matrix are separately invariant under two different subgroups of a larger symmetry group $G$ \cite{deAdelhartToorop:2011re}. The subgroups are usually kept small, whereas in \cite{Holthausen:2012wt} groups $G$ of order up to 1000 have been investigated. It was shown that the possible lepton mixing patterns then depend on how the two subgroups are embedded within $G$. Another approach is based on a Lagrangian with mass terms constructed with lepton, Higgs and scalar flavon fields. Group-invariance of the terms is looked for by systematic probing all plausible models and representation assignments \cite{tribimaximalsmall}; dynamic parameters (VEVs) affect the predicted mixing angles \cite{Lam:2011ag} \cite{Lam:2012ga}. 

Explicit remnant flavor symmetry breaking can be avoided in the presence of Higgs fields that transform under $G$. Two-Higgs doublet models \cite{BRANCO20121} are obvious candidates to investigate that principle. In \cite{Chaber:2018cbi} and \cite{sym12010156} non-Abelian groups have been identified providing lepton masses and a mixing matrix, but these were not in agreement with experiment. Although the implied lepton masses of the model were found nondegenerate, the implied mixing matrices appeared to be always monomial. However, the results are different when one more Higgs doublet is added, as will be outlined below. For an overview of the work on lepton flavor symmetry, we refer to \cite{Feruglio:2019ybq}, \cite{King:2013eh}.

We propose to study a model in which the SM is extended with two $SU(2)$ Higgs doublets \cite{Grossman:1994jb, Keus:2013hya, ChuliCentelles:2022ogma}. The left-handed lepton doublets, the right-handed charged leptons, the right-handed neutrinos as well as the three Higgs doublets themselves are modelled as flavor triplets, each associated with a unitary three-dimensional irreducible representation of a discrete group isomorphic to a subgroup of $U(3)$. Our three-Higgs doublet model (3HDM) contains no flavons or other additional fields. The model allows the treatment of neutrinos as Dirac particles or as Majorana particles. Since the Higgs doublets form a flavor vector transforming under $G$, the mass-squared matrices are not affected by Schur's first lemma: the masses can be nondegenerate and the neutrino mixing nontrivial, even when the assigned 3D irreps are inequivalent and $G$ is non-Abelian. None of the flavor groups with order $\mid G \mid < 1035$ can satisfactorily reproduce the experimental data of the lepton sector in its entirety; only partial symmetries occur. A few groups turn out to be a symmetry of nontrivial Majorana neutrino mixing; the smallest is $\Delta(96)$. This symmetry favours inverse ordering of the neutrinos with masses between 0,013 and 0,05 eV. 

In the following section, the 3HDM and its group-invariance is defined. Section III outlines the (almost fully automated) process of detecting flavor symmetry, and its implications for the lepton masses and mixing angles. In section IV, the results of a scan of groups with $\mid G \mid < 1035$ are presented. Details on the predicted relations among the neutrino mixing angles and lepton masses are provided in Section V. In section VI we summarize the results and discuss directions to modify and generalize the 3HDM.

\section{The Three-Higgs Doublet Model (3HDM)}
In the models which we consider, the unitary symmetry transformations for the fields will not affect most of the terms in the Lagrangian. Although it is known that the Higgs scalar potential \cite{Ivanov_2013} \cite{Ivanov_2013a} are not automatically invariant we solely consider the Yukawa interaction for charged leptons $\mathcal{L}^l$, for Dirac neutrinos $\mathcal{L}^{\nu}$ and for Majorana neutrinos $\mathcal{L}^M$, using an effective dimension-five operator for the latter \cite{Weinberg:1979sa}:
\begin{eqnarray} \label{LAG1}
\mathcal{L}^l = && - (h^l_i)_{\alpha \beta} \overline{L}_{\alpha L} \tilde{\Phi}_i l_{\beta R} + \text{H.c.,}\label{Yukawa1} \\
\mathcal{L}^{\nu} =  && - (h^\nu_i)_{\alpha \beta} \overline{L}_{\alpha L} {\Phi}_i \nu_{\beta R}+ \text{H.c.,} \\
\mathcal{L}^M =  && -\frac{g}{M} (h^M_{i j})_{\alpha \beta}(\overline{L}_{\alpha L} \Phi_i)(\Phi^T_j L^c_{\beta R})+ \text{H.c.,}
\end{eqnarray}
where summation over the lepton flavors $\alpha, \beta = \text{e}, \mu, \tau$ and over the Higgs flavors $i,j = 1,2,3$ is understood. $h^l_i$, $h^\nu_i$ and $h^M_{ij}$ are three-dimensional Yukawa matrices. $\Phi_i$ is an $SU$(2) Higgs doublet; $\tilde{\Phi_i}= i\sigma_2\Phi_i^*$. The $L_{\alpha L}=(\nu_{\alpha L},l_{\alpha L})^T$ are lepton doublets, and
$\bar{L}$ and $L^c$ denote the adjoint and charge-conjugated lepton doublets, respectively.
Due to EWSB, the three mass matrices $M^l$, $M^{\nu}$ and $M^M$ are defined by the mass Lagrangian terms:
\begin{eqnarray} \label{LAG2}
\mathcal{L}^l_{mass} = && - \overline{l}_LM^ll_R  + \text{H.c.}\\
\mathcal{L}^\nu_{mass} = && - \overline{\nu_L}M^{\nu}\nu_R  + \text{H.c.} \\ 
\mathcal{L}^M_{mass} = && - \frac{1}{2}\overline{\nu_L}M^M\nu_L^c + \text{H.c.}
\end{eqnarray}
where $l_L$, $l_R$, $\nu_L$, $\nu_R$ are flavor vectors for the left/right-handed charged leptons and neutrinos, respectively. Each mass matrix is linearly composed of three Yukawa matrices using the vacuum expectation values (VEVs) $v_i$ obtained from $\Phi_i$:
\begin{eqnarray}
M^l && =-\frac{1}{\sqrt2} v^*_ih^l_i \label{mass1}\\
M^{\nu} && =\frac{1}{\sqrt2}v_ih^{\nu}_i  \label{mass2}\\
M^M && =\frac{g}{M} v_iv_jh^M_{ij} \label{mass3}.
\end{eqnarray}
$l_L$, $l_R$, $\nu_L$, $\nu_R$ and $\Phi=(\Phi_1,\Phi_2,\Phi_3)^T$ will each be assigned a three-dimensional irreducible representation of some finite group $G$. All representation matrices should be unitary in order to conserve the total lepton number and to ensure that $\sqrt{\Sigma |v_i|^2}=(\sqrt{2}G_F)^{-1/2}$=246 GeV, where $G_F$ is the Fermi coupling constant.

The first step in this study is to identify distinct groups $G$, isomorphic to a $U(3)$ subgroup, that have one or more three-dimensional irreducible representations, and assign them to flavor vectors so that the Yukawa terms in Eq. (1) and either (2) or (3) remain $G$-invariant. To this end it is determined which of the three terms
\begin{eqnarray} \label{terms1}
&& \overline{L}_L A_L^\dagger\, (A_\Phi^* \tilde{\Phi})_i h^l_i\, A_{l R} \, l_R , \\
&& \overline{L}_L A_L^\dagger\, (A_\Phi \Phi)_i h^\nu_i\, A_{\nu R} \, \nu_R , \label{terms2}\\
&& \overline{L}_L A_L^\dagger\,(A_\Phi \Phi)_i (A_\Phi \Phi'^T)_j) h^M_{ij} \, A^*_L \, L^c_R, \label{terms3}
\end{eqnarray}
(if any) remain unaffected by the simultaneous matrix operators $A_L(g), A_{lR}(g), A_{\nu R}(g)$ and $A_\Phi(g)$ for all $g$ in $G$. The matrix operators are defined by the representations acting on the flavor vectors.
$\Phi'^T=(\Phi_1^T,\Phi_2^T,\Phi_3^T)^T$.

Since the transformations are unitary, the kinetic part of the Lagrangian will be automatically invariant. It can be assumed that also the Higgs potential is unaffected \cite{Chaber:2018cbi}, which justifies our analysis of the Yukawa sector in isolation.

We aim to find symmetry groups leaving both expressions (10) and (11) (and hence $\mathcal{L}^l+\mathcal{L}^{\nu}$) invariant, as we will then be able to derive the implications of the 3HDM on the lepton masses and neutrino mixing angles, in case the neutrinos are Dirac particles. Likewise we look for groups leaving both (10) and (12) (hence $\mathcal{L}^l + \mathcal{L}^M$) invariant, when neutrinos have Majorana nature.

\section{Solving the invariance equations}
At first we treat the three Yukawa terms separately. $G$-invariance of $\mathcal{L}^l$, see Eqs.\eqref{Yukawa1} and \eqref{terms1}, is achieved if and only if
\begin{equation}\label{kron1}
    ((A_\Phi(g))^\dagger\,\otimes\,(A_L(g))^\dagger\,\otimes\,
    (A_{lR}(g))^T)\,h^l\,=\,h^l,
\end{equation}
as demonstrated in \cite{ludl}. The Kronecker product gives a 27$\times$27 matrix, and $h^l$ is the 27-dimensional vector built from the Yukawa matrices $h^l_1, h^l_2$ and $h^l_3$, row-wise. If $h^l$ is an invariant eigenvector satisfying Eq.\eqref{kron1} for all $g \in G$ then $\mathcal{L}^l$ is $G$-invariant. It can be proven that it is sufficient to test the generators of $G$ instead of all $g\in G$. The invariance equations for the terms $\mathcal{L}^\nu$ and $\mathcal{L}^M$ are
\begin{eqnarray}
   ((A_\Phi)^T\otimes(A_L)^\dagger\otimes
    (A_{\nu R})^T) h^\nu && =  h^\nu \label{kron2}\\
    ((A_\Phi)^T\otimes(A_\Phi)^T\otimes(A_L)^\dagger
    \otimes(A_L)^\dagger) h^M && =  h^M \label{kron3},
\end{eqnarray}
dropping the $(g)$-argument for clearness. $h^\nu$ is a 27-dimensional invariant eigenvector built from the entries of the three $h^\nu_i$ matrices, row-wise. The 81-dimensional vector $h^M$ contains the entries of $h^M_{11}$, row-wise, followed by the entries of $h^M_{12}$, $h^M_{13}$, $h^M_{21}$ etc.

Explicit solutions of equations (13), (14) or (15) are computationally demanding for groups of a high order, as they may have a large number of representations and thus give rise to many combinations of the representations assigned to the flavor vectors.  However, we will describe here that based on the character table of $G$, we can filter away the irrelevant representation assignments, thus reducing the computational task enormously.

Eqs.\eqref{kron1}, \eqref{kron2} and \eqref{kron3} turn out to be equivalent to the (Clebsch-Gordan) tensor product decomposition equations \cite{ludl}
\begin{eqnarray}
    A_L \otimes A_{lR}^* \, && =  A_\Phi^* \oplus ...\label{decomp1} \\
    A_L \otimes A_{\nu R}^* \, && =  A_\Phi \oplus ... \label{decomp2} \\
    A_\Phi^* \otimes A_L \otimes A_L \, && =  A_\Phi \oplus ... \label{decomp3}
\end{eqnarray}
Eqs.\eqref{decomp1} (or \eqref{decomp2}) are fulfilled if and only if the 9-dimensional tensor product operator allows a decomposition containing at least one three-dimensional matrix operator. Eq.\eqref{decomp3} requires that the 27-dimensional tensor product operator contains at least one three-dimensional matrix operator. All three cases can be verified by reading off the group's character table, as detailed in the next section. From the solutions $h^l, h^\nu$ or $h^M$, the mass matrices are constructed enabling us to pin down the lepton mass ratios and/or the neutrino mixing angles to certain values or intervals.

\section{Finding the $G$-invariant Lagrangians for $\mid G \mid < 1035$}
A candidate group $G$ rendering a mass term $G$-invariant must have at least one faithful three-dimensional representation (otherwise $G$ would not be isomorphic to a subgroup of $U(3)$). Unfaithful representations are also included in this analysis, despite the abundant repetition of representation assignments that can be expected in groups containing $G$. Our precise selection criterion is that at least one of the irreducible representations assigned to a mass term is faithful. Out of the $U(3)$ subgroups with $\mid G \mid < 1035$, 749 groups provide solutions to Eq.\eqref{kron1} and, in these cases, also to Eq. \eqref{kron2}. 216 groups provide solutions to Eq.\eqref{kron3}.

The selection and processing of groups is fully automated using the computer-algebra system GAP \cite{GAP4}. To determine which representation assignments would solve a particular invariance equation it is sufficient to observe the group's character table, which is readily provided by GAP.
The character of $g\in G$ in representation $A$, denoted $\chi^A (g)$, is defined as the trace of matrix $A(g)$. The mapping $\chi^A$ is called the character of $A$. Let $A$ and $B$ be representations of $G$, then
\begin{equation}
\langle \chi^A,\chi^B \rangle := \frac{1}{|G|} \sum_{g \in G}\chi^A (g)^\star \chi^B (g),
\end{equation}
defines the inner product of $\chi^A$ and $\chi^B$. It can be proven that $\chi ^{A \otimes B}(g) = \chi^A(g) \chi^B(g)$ for all $g \in G$. Let also $C$ be an irreducible representation of $G$. Then $\langle \chi^{A \otimes B},\chi^C \rangle$  is the multiplicity of $C$ occurring in the tensor product decomposition of $A \otimes B$. For the decomposition in Eq. \eqref{decomp1}, $\langle \chi^{A_L \otimes A_{lR}^*},\chi^{A_\Phi^*} \rangle$ can take the values 0 to 3. This is the number of linearly independent solutions $h^l$ to Eq. \eqref{kron1}. The inner products can be directly deduced from the character table of $G$, and thus prior to the actual generation of the representation matrices themselves and without explicitly solving Eq. \eqref{kron1}. For brevity let us denote the representations appearing in Eq. \eqref{decomp1} as $A$, $B$ and $C$, respectively. Then, if $A$, $B$ and $C$ are irreducible, Eq. \eqref{kron1} has a nontrivial solution if and only if
$n_C :=\langle \chi^{A \otimes B^\star},\chi^{C^\star} \rangle > 0$.

In the selection procedure the representation triplet $(A,B,C)$ is accepted only if $n_C=1$; as a trade-off regarding computational load, we disregard multidimensional solutions ($n_C>1$). We find over 6 million accepted triplets. In the following step, the explicit three-dimensional matrix representations (denoted $\mathbf{3}_A,\mathbf{3}_B$ and $\mathbf{3}_C$) of $A$, $B$ and $C$ are obtained using the Repsn package of GAP.
\cite{Repsn}
The Kronecker product Eq.\eqref{kron1} is set up for each generator of $G$, and solved for $h^l$, using the BaseFixedSpace function of GAP. The total number of inequivalent vectors $h^l$ from the group scan is 2130 (a set of inequivalent vectors contains no colinear pairs; colinear solutions would imply mass matrices differing by a constant only).  For the Dirac neutrino term, with $n_C=\langle \chi^{A \otimes B^\star},\chi^C \rangle$ we find the same number of solutions. For the Majorana term, only two characters are involved; let us denote them $\chi^A$ and $\chi^C$. The character inner product $\langle \chi^{C^\star \otimes A \otimes A}, \chi^C \rangle$ can take the values 0 to 9, equal to the dimension of the solution space. Again, only solutions with inner product equal to one are accepted. We find 70 inequivalent solutions $h^M$.
Using Eqs. \eqref{mass1}, \eqref{mass2}, \eqref{mass3} we obtain the mass matrices as functions of $v_i$. For this calculation and the subsequent (mostly numerical) computations, we use the Mathematica package from Wolfram \cite{Mathematica}.

If the triplet of representations $(\mathbf{3}_A,\mathbf{3}_B,\mathbf{3}_C)$ yields a $G$-invariant charged-lepton term, then this triplet can also render the Dirac neutrino term $G$-invariant.
Can any charged-lepton triplet $(\mathbf{3}_A,\mathbf{3}_B,\mathbf{3}_C)$ be combined with any Dirac neutrino triplet $(\mathbf{3}_D,\mathbf{3}_E,\mathbf{3}_F)$ to obtain a $G$-invariant charged-current lepton interaction term? The answer is no; we must require $\mathbf{3}_D=\mathbf{3}_A$ because the left-handed charged lepton and the left-handed Dirac neutrino are contained in the same $SU(2)_L$ doublet and hence transform equally. Furthermore, we require $\mathbf{3}_F=\mathbf{3}_C$ since the two states in a Higgs doublet respect $SU(2)_L$ symmetry and thus differ by complex conjugation. So we only consider representation assignments of the form $((\mathbf{3}_A,\mathbf{3}_B,\mathbf{3}_C),(\mathbf{3}_A,\mathbf{3}_D,\mathbf{3}_C))$. The simultaneous solution of Eqs.\eqref{kron1} and \eqref{kron2} gives mass matrices $M^l$ and $M^\nu$ that define the PMNS matrix.
In the case where neutrinos are Majorana particles, $G$-invariance of the charged-current interaction term requires assignments of the form
$((\mathbf{3}_A,\mathbf{3}_B,\mathbf{3}_C),(\mathbf{3}_A,\mathbf{3}_C))$.

For Dirac neutrinos the PMNS matrix is calculated with the two unitary matrices $U_l$ and $U_\nu$ that diagonalize the mass-squared matrices for charged leptons and for neutrinos, respectively:
\begin{equation}
U_l^\dagger(M^l M^{l \dagger}) U_l = (M_d^l)^2, \, \, \,
U_\nu ^\dagger(M^\nu M^{\nu \dagger}) U_\nu = (M_d^\nu)^2,
\end{equation}
where the subscript $d$ denotes the diagonal matrix. In the case of Majorana neutrinos their mass matrix is symmetric, and is diagonalized using one unitary matrix $U_\nu$:
\begin{equation}
U_\nu^T M^M U_\nu= M^M_d.
\end{equation}
Independently of the neutrinos' nature the PMNS matrix is given by $U_l^\dagger U_\nu$.
\begin{table*}
\centering
\caption{\label{tab:table1}Comparison of the 3HDM predictions to the experimental data,
for selected groups.}
\begin{tabular}{lclcccc}
& & &                                      Dirac &Maj.   &Dirac                &Maj. \\
GAP-ID & Structure & 3D Irreps (Faithful) & mass & mass & mixing $\chi^2 /4$ & mixing $\chi^2 /4$ \\
 \hline
$[12,3]$&$A_4$&\hspace{30pt}1 (1)&-&-&-&-\\
$[21,1]$&$T_7$&\hspace{30pt}2 (2)&a&-&d&-\\
$[24,12]$&$S_4$&\hspace{30pt}2 (2)&b,c&-&5.3&-\\
$[39,1]$&$T_13$&\hspace{30pt}4 (4)&a&-&d&-\\
$[48,3]$&$\Delta(48)$&\hspace{30pt}5 (4)&a&a&d&d\\
$[48,30]$&$A4:C4$&\hspace{30pt}4 (2)&b&-&d&- \\
$[48,48]$&$C2\times S4$&\hspace{30pt}4 (2)&b,c&-&5.3&- \\
$[60,5]$&$A5$&\hspace{30pt}2 (2)&b&-&d&- \\
$[72,42]$&$C3\times S4$&\hspace{30pt}6 (4)&b,c&-&5.3&-\\
$[84,11]$&&\hspace{30pt}9 (6)&a&a&d&d\\
$[96,64]$&$\Delta(96)$&\hspace{30pt}6 (4)&a,b,c&a,c&d&0.0\\
$[96,68]$&&\hspace{30pt}10(4)&a&a&d&d\\
$[96,186]$&$C4\times A4$&\hspace{30pt}8 (4)&b.c&-&5.3&-\\
$[108,15]$&$\Sigma(36\times3)$&\hspace{30pt}8 (8)&b,c&-&-&-\\
$[120,37]$&$C5\times S4$&\hspace{30pt}10 (8)&b,c&-&5.3&- \\
$[150,5]$&$\Delta(150)$&\hspace{30pt}8 (8)&a,b,c&a,b,c&d&0.0\\
$[192,182]$&$\Delta(96,2)$&\hspace{30pt}12 (4)&a,b,c&a,b&d&0.0\\
$[192,944]$&$C2\times \Delta(96)$&\hspace{30pt}12 (4)&a,b,c&a,c&d&0.0 \\
$[243,19]$&$Z^{\prime\prime}(3,3)$&\hspace{30pt}24 (18)&a,c&-&150&-\\
$[432,239]$&$\Pi(1,2)$&\hspace{30pt}16 (8)&b&a&d&d\\
$[729,63]$&$Z^{\prime\prime}(3,4)$&\hspace{30pt}72 (54)&a,c&-&150&-\\
$[864,675]$&$\Pi(1,3)$&\hspace{30pt}32 (16)&b&a&d&d\\
\end{tabular}
\end{table*}

For a given group we can identify group-invariant mass terms and derive masses and mixing angles as functions of the VEVs for each particular representation assignment. From group theory alone, we can neither determine the absolute scale of the Higgs couplings, so we have (at most) $v_2/v_1$ and $v_3/v_1$ as four free real parameters, and we can at most determine mass ratios $m_i/m_j$. It will be only possible to determine the PMNS matrix up to permutation of rows or columns and (for the Majorana case) up to a phase for two rows.

The results from selected groups are presented in Table 1. In its first column the GAP-ID is the identifier supplied by the SmallGroups library of GAP.
\cite{SmallGroup}.
The first index equals the group's order, the second distinguishes between the non-isomorphic groups of that order. The second column of the table shows the group structure (when informative). The number of three-dimensional irreducible representations and of faithful ones, are listed in column 3. The following cases are distinguished in the "Dirac Mass" and "Maj. Mass" columns. A minus sign "-" means that the tensor product could not be decomposed for any of the representation assignments otained for the group. "a" means that mass ratios consistent with the experimental values can be obtained. "b" indicates one, two or three vanishing or degenerate masses. "c" indicates an upper bound for $m_3/m_2$ while $m_2/m_1$ and $m_3/m_1$ can obtain any positive value, as functions of the  $v_i$. Multiple tokens mean different results for different representation assignments for the group. The columns labeled "Dirac mixing" and "Maj. Mixing" indicate how close a predicted PMNS matrix gets to the experimental data, expressed as $\chi^2 /4$, the average of the deviations between calculated and experimental values of the four quantities: $\sin^2\Theta_{12}$, $\sin^2\Theta_{23}$, $\sin^2\Theta_{13}$. "d" signifies that only monomial PMNS matrices are found. $\chi^2$ is derived from a simplified extraction of the experimental neutrino oscillation parameters, see Table II. $\chi^2$ is the smallest value found by equidistant numerical sampling in a region of four-dimensional $(v_2/v_1,v_3/v_1)$-space. There is no proof that we find the global minimum.

We find no group representations implying the lepton masses and mixing angles to be simultaneously consistent with the experimental data. Group $T_7$ is the smallest group with invariant $\mathcal{L}^l + \mathcal{L}^\nu$ compatible either with the experimental charged-lepton masses or with the experimental neutrino mass data, with different VEVs for either case. Group $\Delta(48)$ is the smallest group with invariant $\mathcal{L}^M$ compatible with the experimental neutrino mass data. For Majorana neutrinos, the smallest group compatible with the  PMNS data is $\Delta(96)$, (that is $\mathcal{L}^l + \mathcal{L}^M$ invariant). For Dirac neutrinos we find no group exactly compatible with the PMNS data. $S_4$ (and groups containing it) comes closest, with $\chi^2=5.3$.
In the next section, further details of the $\Delta(96)$ and $S_4$ are presented in subsections A and B, respectively. Other groups generating specific solutions are described in subsection C.

\section{$G$-invariant masses and mixing}
\subsection{\label{sec:5.1}Group $\Delta(96)$}
$\Delta(96)$ is a symmetry group of lepton mixing if we assume the neutrinos to be Majorana particles, and apply representation assignment $((\mathbf{3}_1,\mathbf{3}_3,\mathbf{3}_6),
\,(\mathbf{3}_1,\mathbf{3}_6))$. Here the representations $\mathbf{3}_i$ are from those provided by Repsn (six 3D irreps in total). $\mathbf{3}_1$ and $\mathbf{3}_2$ are the two unfaithful representations and $\mathbf{3}_3 \dots \mathbf{3}_6$ are faithful, with $\mathbf{3}_4=\mathbf{3}_3^\star$ and $\mathbf{3}_6=\mathbf{3}_5^\star$. The mass matrices obtained from $h^l$ and $h^M$ using Eqs.\eqref{mass1} and \eqref{mass3} are of the form
\begin{equation}\label{massdelta96}
\begin{array}{cc}
M^l =-\frac{c_l}{\sqrt2}
\begin{pmatrix}
0       &   0        &      v_3^\star     \\
v_1^\star  &   0        &      0     \\
0       &   v_2^\star        &      0     
\end{pmatrix},
 M^M = -\frac{c_Mg}{M}
\begin{pmatrix}
0       &   v_2^2        &      v_1^2     \\
v_2^2       &   0        &      v_3^2     \\
v_1^2       &   v_3^2        &      0     
\end{pmatrix},
\end{array}  
\end{equation}
where $c_l$ and $c_M$ are arbitrary constants inherent to the two $h$ matrices.
From these follow the two mass ratios of the charged leptons, the two mass ratios of the neutrinos (or two mass-squared differences up to a common factor), and the four neutrino mixing angles. All 8 quantities are functions of the VEVs. We search for values $v_i$ yielding the mass ratios and/or mixing parameters in agreement with experimental data. The four calculated mixing angles are consistent with experimental data with high accuracy for determined regions in the four-dimensional search space. The search is implemented numerically by an equidistant sampling of $(|v_2/v_1|$, arg$(v_2/v_1), |v_3/v_1|$, arg$(v_3/v_1)$. The search box is, at present, limited to [0,4000] for the $|v_i/v_1|$ ratios. We find multiple choices of the VEVs fitting the mixing data with $\chi^2 < 10^{-3}$. In general it will be possible to detect only a subset of solutions yielding consistent mixing angles. For the charged leptons the mass ratios are equal to $|v_i/v_1|$, as obtained from $M^l$ in Eq.\eqref{massdelta96}. The charged lepton mass ratios found until present remain far below the experimental values by approximately a factor 2 for the muon to 35 for the tau. We do obtain fits simultaneously to the Majorana neutrino mass-squared data \textit{and} to the mixing angles. The best fit ($\chi^2=0.3$) is achieved with modulus of eigenvalues of $M^M$ equal to (0.217, 0.867, 0.879). ($\chi^2$ here is averaged over six quantities: two neutrino mass-squared differences and four mixing angles). Scaled to the experimental data it corresponds to (inverted ordered) neutrino masses $m_1=12.70\times10^{-3}$eV, $m_2=50.78\times10^{-3}$eV, $m_3=51.49\times10^{-3}$eV. This fit is obtained with: $|v_2/v_1|=0.31$, arg$(v_2/v_1)=0.81$,$(|v_3/v_1|=0.93$, arg$(v_3/v_1)=2.15$.

Let us point out some remarks on the uniqueness of this result. Numerical explorations suggest finite, possibly disconnected, regions in VEV space with $\chi^2<0.5$, so there are multiple choices of the $v_i$ yielding neutrino mixing and masses consistent with experimental data. Besides $((\mathbf{3}_1,\mathbf{3}_3,\mathbf{3}_6),
\,(\mathbf{3}_1,\mathbf{3}_6))$ there is another valid representation assignment:  $((\mathbf{3}_3,\mathbf{3}_4,\mathbf{3}_2),
\,(\mathbf{3}_3,\mathbf{3}_2))$ that also renders the  PMNS matrix invariant, which however is monomial, ruling out any flavor mixing. It turns out that there are no further representation assignments inequivalent to the two just described; two representation assignments are equivalent if their sets of solution vectors $(h^l,h^M)$ are linearly dependent. We note that the $\Delta(96)$-symmetry involves the unfaithful representation $\mathbf{3}_1$. Out of the $6^3=216$ permutations, 40 representation assignments for the charged lepton mass term can realize the tensor product decomposition (Eq.\eqref{decomp1}) and hence define solution vectors $h^l$. If we remove vectors such that no two vectors differ from each other by a constant, only 5 inequivalent vectors are left. The same numbers come for the Dirac neutrino mass term
(Eq.\eqref{decomp2}). For the Majorana mass term (Eq.\eqref{decomp3}) there are 24 valid representation assignments  of which 3 are inequivalent. Finally, for the neutrino mixing term we find 16 valid representation assignments, 8 of which are equivalent to $((\mathbf{3}_1,\mathbf{3}_3,\mathbf{3}_6),\,(\mathbf{3}_1,\mathbf{3}_6))$.
So we conclude that this representation assignment, as a provider for $\Delta(96)$ mixing symmetry, is not unique and, consequently the 8 pairs of invariant eigenvectors $(h^l,h^M)$ are mutually equivalent. Similar pairs also appear in higher-order groups, such as $\Delta(150)$ and $C2\times \Delta(96)$.
\begin{table}
\centering
\caption{\label{tab:table2}
Experimental oscillation parameters. Extracted from \cite{deSalas:2020pgw}.
}
\begin{tabular}{lccc}
Parameter&Value&$\sigma$ \\
\hline
$\Delta m_{21}^2 (10^{-5}$eV$^2)$ & 7.5 & 0.18 \\
$\Delta m_{32}^2 (10^{-5}$eV$^2$) & 249 & 3.5 \\
$sin^2\Theta_{12}$ & 0.310 & 0.015 \\
$sin^2\Theta_{23}$ & 0.566 & 0.025 \\
$sin^2\Theta_{13}$ & 0.0224 & 0.0007\\
$\delta_{CP}$ (\textdegree) & 250 & 35 \\
\end{tabular}
\end{table}

\subsection{Group $S_4$}
$S_4$ is the smallest group allowing nontrivial flavor mixing when neutrinos are Dirac particles. The two representation assignments providing this symmetry are 
$((\mathbf{3}_1,\mathbf{3}_1,\mathbf{3}_1),\,(\mathbf{3}_1,\mathbf{3}_1,\mathbf{3}_1))$ and
$((\mathbf{3}_1,\mathbf{3}_1,\mathbf{3}_2),\,(\mathbf{3}_1,\mathbf{3}_1,\mathbf{3}_2))$, where $\mathbf{3}_1$ and $\mathbf{3}_2$ are the two inequivalent three-dimensional representations of $S_4$. The first representation assignment leads to anti-symmetric mass matrices and trivial mixing. The second implies the mass matrices (using Eqs. (7) and (8)):
\begin{equation}
\begin{array}{cc}
M^l = -\frac{c_l}{\sqrt2}
\begin{pmatrix}
0       &   v_3^\star        &      v_2^\star     \\
v_3^\star  &   0        &      v_1^\star     \\
v_2^\star       &   v_1^\star        &      0     
\end{pmatrix},
 M^{\nu} = -\frac{c_\nu}{c_l} M^{l*},
\end{array}  
\end{equation}
where $c_\nu$ is an arbitrary constant. We obtain fits to the neutrino mixing data with $\chi^2$ = 5.3. The deviation is mainly due to $\sin^2\Theta_{23}$ and $\sin^2\Theta_{13}$, which end up approximately $3\sigma$ larger than the observed values. 
The same results are found in higher-order groups, such as $C_i\times S_4$ and $C_4\times A_4$.

The predicted mass ratios $m_\nu/m_e$ and $m_\tau/m_e$ are too small and do not come close to the actual charged lepton mass ratios. However, the Dirac neutrino mass ratios are consistent with the experimental data. A fit to the mass term alone - not combined with neutrino mixing - has $\chi^2<0.01$ and gives (inverted ordered) neutrino masses:
$(m_1, m_2, m_3)$ = (0.733, 49.17, 49.91)$\times10^{-3}$eV; the combined fit of Dirac neutrino masses and mixing has $\chi^2=8$.

\subsection{Other groups}
Group $\Sigma(36\times3)$ is the lowest-order group generating a Dirac mass matrix with 9 non-zero entries. 96 representation assignments generate a Dirac mass matrix (the set of 96 representations form 5 mutually inequivalent subsets), none being compatible with the observed masses. None of the representation assignments gives an invariant interaction term.

Further new types of mass matrices show up in group $Z^{\prime\prime}(3,3)$. For this group we find 1134 representation assignments (forming 109 inequivalent subsets), each giving an invariant Dirac mass matrix, and 2592 assignments (forming 74 inequivalent subsets) generate an invariant PMNS matrix for Dirac neutrinos. It turns out that when an unfaithful representation occurs in the assignment, a $\chi^2\approx 150$ fit to the oscillation data can be obtained, whereas for all other representation assignments the PMNS matrix is monomial. The group $Z^{\prime\prime}(3,4)$ generates similar results, with further new types of Dirac matrices.

$\Pi(1,2)$ generates a new type of Majorana mass matrix, giving neutrino masses with normal ordering: $(m_1,m_2,m_3)=(17.3, 19.3, 52.81)\times 10^{-3}$eV. The charged lepton mass matrix is anti-symmetric, disabling nontrivial neutrino mixing. Another new type of Majorana mass matrix is generated by $\Pi(1,3)$, providing results similar to those of $\Pi(1,2)$.

\section{Conclusions}
In the 3HDM model, we have searched for a discrete flavour symmetry that would predict the hierarchy of charged leptons and their mixing. The investigation comprised all discrete groups having three-dimensional irreducible representations up to the order of 1035. With the applied representation selection criteria, none of the studied groups is a symmetry of the entire lepton sector. There exist symmetries that separately predict the masses of charged leptons, the masses of neutrinos, and/or the elements of the PMNS mixing matrix, as can be expected for a model with four free parameters.

The most noticeable results are obtained with $\Delta(96)$ and $S_4$.
The smallest group compatible with the neutrino mixing data is $\Delta(96)$, when assuming that neutrinos have Majorana nature. For the same parameters $v_i$ the predicted neutrino masses are consistent with the experimental data as well. The predicted charged lepton masses are then far off the experimental values.

$S_4$ is the smallest group approximately compatible with the experimental neutrino mass \textit{and} mixing data, in case the neutrinos are Dirac particles. The $v_i$ producing that fit imply masses of the charged leptons which are in disagreement with experiment. 

It should be noted that we limited our search for proper
discrete symmetry to groups with irreducible three-dimensional faithful
representations, and we searched for non-degenerate eigenvectors only,
constructed from Yukawa matrices for leptons and neutrinos. In this
context, we can say that the problem of mass for charged leptons
and neutrinos and their mixing in charged currents still awaits a solution.

There are several possible improvements and extensions to the described methods.
The present results are based on numerical sampling in four-dimensional
VEV-space, where the choice of search interval and the sampling density is
limited for practical reasons. The analysis would be highly enhanced
when analytic expressions for the eigenvalues of mass matrices are used
to find bounds on physical quantities.
The restrictions we made on the representation assignments are mostly
for practical reasons, to limit the computational work. It is therefore
worthwhile to explore the inclusion of unfaithful and reducible
representations. Also, the requirement that the tensor product
decomposition is unique could be relaxed, allowing multi-dimensional
solutions.\\
\vskip 0.3cm
\hskip -0.5cm
{\large \bf Acknowledgments}
\vskip 0.2cm
This work has been supported in part by the Polish National Science Center (NCN) under grant 2020/37/B/ST2/02371 and the Research Excellence Initiative of the University of Silesia in Katowice. We thank Jacek Holeczek for his help. We are very grateful to the GAP Support Group for their advising.
\printbibliography
\end{document}